\documentclass[sigconf,screen,nonacm]{acmart}

\renewcommand\footnotetextcopyrightpermission[1]{}
\AtBeginDocument{%
  }

\begin{document}

\title[Knowledge-Based Pull Requests]{Knowledge-Based Pull Requests: A Trusted Workflow for Agent-Mediated Knowledge Collaboration}

\hypersetup{
  pdftitle={Knowledge-Based Pull Requests: A Trusted Workflow for Agent-Mediated Knowledge Collaboration},
  pdfauthor={Xinyu Zhang and Weiwei Sun},
  pdfsubject={Knowledge-Based Pull Requests and agent-mediated software collaboration across trust boundaries}
}

\author{Xinyu Zhang}
\email{zhangxinyu@fudan.edu.cn}
\affiliation{%
  \institution{Fudan University}
  \city{Shanghai}
  \country{China}}

\author{Weiwei Sun}
\email{wwsun@fudan.edu.cn}
\affiliation{%
  \institution{Fudan University}
  \city{Shanghai}
  \country{China}}

\renewcommand{\shortauthors}{Zhang and Sun}

\begin{abstract}
AI coding agents are changing the bottleneck in software collaboration:
code is increasingly cheap, while understanding intent, negotiating scope, and
governing long-term project responsibility remain costly. This paper proposes
\emph{Knowledge-Based Pull Requests} (KPR), a trusted workflow for
agent-mediated software collaboration across trust boundaries, including open
source, enterprise, vendor, contractor, and customer-driven settings. In KPR,
an external collaborator's local code, tests, and cleaned agent interaction
trace are treated as knowledge sources rather than as the default merge
candidate. Agents distill these sources into a human-confirmed knowledge
package and render it into reviewer-facing forms such as design memos, risk
checklists, test plans, or implementation briefs. A project-owned inner trusted
coding agent then regenerates candidate code inside the receiving project's
environment under repository context, engineering conventions, tests, and
security policy. KPR therefore separates two decisions that traditional pull
requests often collapse: whether the knowledge should enter the project, and
whether a particular implementation should be merged. We contribute the KPR
workflow, a candidate artifact schema, a cost-accounting view, a collaboration
gateway architecture, a minimal controlled simulation pilot over seven merged
public pull requests, and an evaluation agenda. The pilot shows that KPR
packages can be instantiated from real PR material and stress-tested under
description ablation, diff ablation, and synthetic poisoned-patch conditions.
We position KPR as an empirically testable workflow: its value depends on
whether auditable extraction, transformation, and project-side regeneration
reduce the cost of understanding and reworking high-context external changes.
\end{abstract}

\ccsdesc[500]{Software and its engineering~Software development process management}
\ccsdesc[300]{Human-centered computing~Collaborative interaction}
\ccsdesc[100]{Computing methodologies~Planning and scheduling}

\keywords{pull requests, coding agents, trusted collaboration, cross-trust-boundary software collaboration, human-agent collaboration, software governance, provenance, code review}

\begin{teaserfigure}
  \centering
  \includegraphics[width=\textwidth]{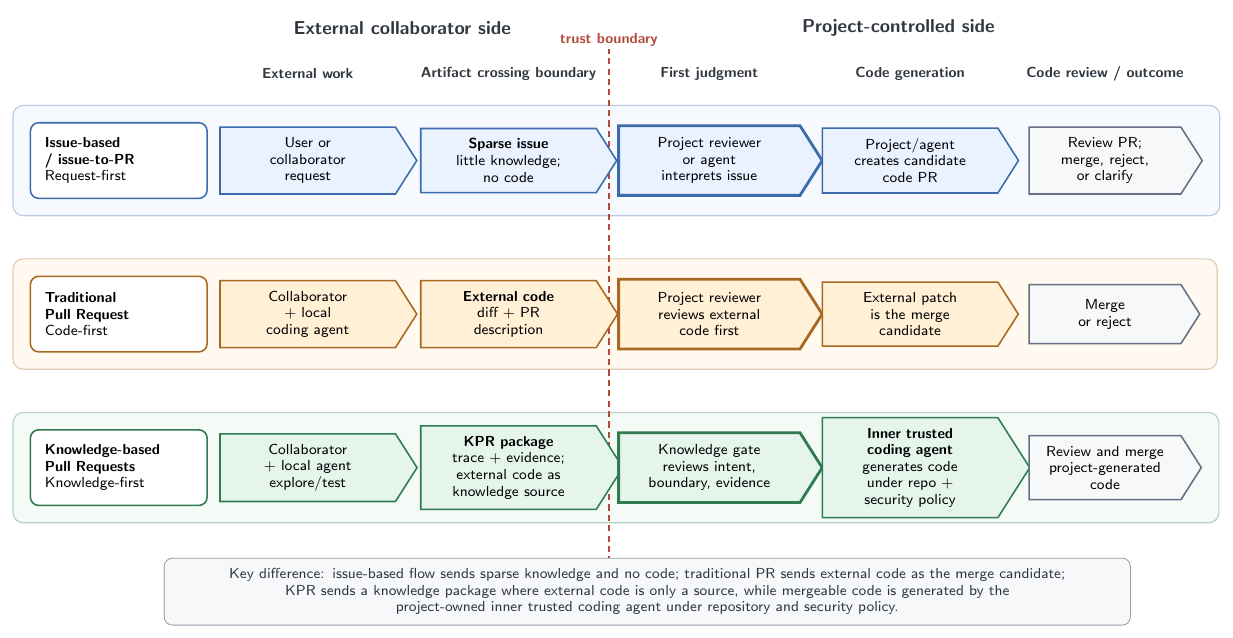}
  \Description{A three-row process overview comparing issue-based,
  traditional pull request, and KPR workflows. The
  issue-based row shows sparse knowledge and no code crossing the trust
  boundary. The traditional pull request row shows external code crossing the
  boundary as the merge candidate. The KPR row shows a
  knowledge package crossing the boundary, with mergeable code generated later
  by a project-owned inner trusted coding agent under repository and security
  policy.}
  \caption{Comparison of issue-based, traditional pull request, and
  KPR workflows. In issue-based flows, sparse
  knowledge and no code cross the boundary. In traditional PRs, external code
  crosses the boundary and becomes the merge candidate. In KPR, external code
  is only one source in a knowledge package; the code that can enter the
  codebase is generated by a project-owned inner trusted coding agent under
  repository and security policy.}
  \label{fig:kpr-overview}
\end{teaserfigure}

\maketitle

\section{Introduction}

The pull request (PR) became the default collaboration interface for modern
open source, and PR-like workflows now shape many forms of software
collaboration: internal platform teams reviewing product-team changes,
contractors delivering patches to a client repository, vendors contributing to
an integration layer, and customer-facing teams forwarding local
customizations to a core product team. These workflows package a proposed code
change together with review, discussion, automation, and an integration
decision. They also assume that the code diff is the central artifact. A
collaborator writes or generates a patch, submits it to a project, and
project-side reviewers judge whether that patch should enter the codebase.

AI coding agents put pressure on that assumption because they can now perform
much of the mechanical and interpretive labor around a change. A collaborator
can ask a local agent to explore an unfamiliar repository, modify many files,
run tests, summarize a diff, and open a plausible PR in a fraction of the time
that a human author would have needed. The project-side bottleneck is therefore
no longer only implementation capacity. It is collaboration burden: project
owners and reviewers must still decide whether the need is real, whether the
proposed design respects project boundaries, which evidence matters, how the
change interacts with internal policy, and what long-term responsibility the
project accepts after merge.

This burden is clearest when software work crosses a trust boundary. In open
source, that boundary separates external contributors from project maintainers.
In closed or enterprise settings, it may separate a vendor from a product team,
a contractor from an internal repository, a customer deployment from the core
platform, or two internal teams with different access rights, compliance
obligations, and architectural knowledge. The shared pattern is the same:
external work may contain useful knowledge, but the receiving project cannot
simply trust the external implementation as the code that should enter the
shared system.

Recent ecosystem signals from public code collaboration platforms make this
tension visible. One major platform provider has published practical guidance
for reviewing agent-generated PRs~\cite{githubAgentPRs2026} and added
repository-level controls to disable PRs, restrict PR creation to
collaborators, and limit concurrent PRs from users without write
access~\cite{githubPRAccess2026,githubPRLimit2026}. Individual projects are
also adjusting their contribution policies. For example, one public
coding-agent project's contribution guide requires issue linkage and design
review for certain kinds of work, and asks contributors to explain changes in
their own words rather than submit long AI-generated
descriptions~\cite{opencodeContributing}. These
changes do not prove that agentic PRs are intrinsically bad. They do show that
review bandwidth and contribution governance are becoming first-class
constraints on public code collaboration platforms. The same constraint appears
less publicly inside organizations, where the receiving team must still
translate external context into internal requirements, policy checks, and
implementation decisions.

Empirical software engineering research points in the same direction. Njoku
et~al. compare 40,214 PRs across 2,807 repositories, including 33,596
agent-authored PRs, and find a socio-technical trade-off: agent-authored PRs
are integrated faster but have lower overall merge rates than human-authored
PRs~\cite{njoku2026codeauthors}. Ehsani et~al. analyze 33k agent-authored PRs
and report that documentation, CI, and build updates merge more successfully
than performance and bug-fix work, while failed agentic PRs tend to touch more
files and fail CI more often~\cite{ehsani2026where}. Peralta et~al. further
argue that merge and rejection labels alone are insufficient: in their manually
inspected rejected cases, only 35.7\% reflect clear agent failures, while many
are driven by workflow constraints or lack observable decision
rationale~\cite{peralta2026why}. Together, these findings suggest that the
problem is not merely patch correctness. It is also whether the collaboration
interface lets project-side developers delegate machine-readable work to agents
while preserving human control over judgment, policy, and merge authority.

This paper proposes \emph{Knowledge-Based Pull Requests} (KPR), a trusted
workflow for agent-mediated software collaboration across trust boundaries. Its
design principle is agent-centered but project-controlled: structured,
auditable, and low-judgment information work should be delegated to agents when
doing so can be checked, while decisions that commit the project to product
direction, architecture, risk, responsibility, credit, or merge remain under
project-side human control. In KPR,
external local code becomes one knowledge source alongside cleaned agent
traces, tests, reproduction evidence, rejected alternatives, and human edits.
Agents distill these sources into a human-confirmed knowledge package and can
transform that package into the review format preferred by the receiving
project. Only after this knowledge passes a project-side gate does a
project-owned inner trusted coding agent regenerate candidate code under
repository context, conventions, tests, and security policy.

The key shift is therefore not ``agents write code.'' Agents already write
code. The key shift is that agents are used to reshape where collaboration work
happens without moving implementation authority outside the receiving project.
In traditional PRs, an external patch crosses the trust boundary and is reviewed
as the merge candidate. In KPR, a knowledge package crosses the boundary, while
mergeable code is generated inside the project-controlled environment.

This paper presents a conceptual research framework and evaluation agenda. It
uses open source PRs as the motivating and empirically visible case, but the
proposed mechanism applies more broadly to software collaboration whenever
external knowledge and code cross into a project-controlled environment. The
enterprise, vendor, contractor, and customer-deployment examples in this paper
should be read as plausible extensions of the same trust-boundary pattern, not
as empirically validated deployment settings. The paper does not claim that KPR
has already reduced review burden in practice. Its contribution is to make the
workflow precise enough to be debated, prototyped, and evaluated. We make four
contributions:

\begin{itemize}
  \item We frame agent-mediated software collaboration as a problem of
  redistributing project-side collaboration burden across trust boundaries, not
  only generating code.
  \item We define the KPR workflow and distinguish it from issue-based
  automation, traditional PRs, design docs, and spec-driven development.
  \item We propose a candidate KPR artifact schema, a cost-accounting view, and
  a collaboration gateway architecture that use agents to summarize traces,
  generate reviewer-facing views, and drive project-side regeneration.
  \item We report a minimal controlled simulation pilot over seven merged public
  PRs and outline an empirical evaluation agenda comparing structured
  PR/issue submissions and KPR packages with project-side regeneration.
\end{itemize}

\section{Background and Related Work}

KPR sits at the intersection of five lines of work: pull-based development and
modern code review, automated agents in software work, agent-authored PRs and
input quality, spec-driven development, and provenance-aware trustworthy
agents. We organize the related work around the collaboration problem rather
than around individual tools. Prior work shows that software collaboration is
not only a code production pipeline: it is also a process of explaining intent,
judging fit, transferring rationale, and governing responsibility across
organizational or project boundaries.

\subsection{Pull Requests as Collaboration and Governance}

Pull-based development already separates implementation effort from integration
authority. Gousios et~al.'s exploratory study shows that project, contributor,
and change factors all influence PR merge decisions and processing
time~\cite{gousios2014pullbased}. Subsequent studies make the two sides of
this workflow explicit: integrators must manage contribution quality,
prioritization, and acceptance at scale~\cite{gousios2015integrator}, while
contributors face process and communication challenges when trying to get
changes accepted~\cite{gousios2016contributor}. KPR preserves this separation,
but changes the artifact that first crosses the boundary from external code to
structured project-relevant knowledge.

Modern code review research further shows that review is not merely defect
detection. Bacchelli and Bird find that review supports knowledge transfer,
team awareness, alternative solution creation, and change
understanding~\cite{bacchelli2013codereview}. Sadowski et~al.'s case study of
code review at Google shows that lightweight tool-based review is an
institutionalized workflow for large-scale engineering practice, not an ad hoc
correctness check~\cite{sadowski2018googlecodereview}. These findings motivate
KPR's central cost claim: for high-context changes, the expensive review work is
often understanding, scoping, and governing the change rather than reading
syntax alone.

Research on rationale and PR descriptions explains why the final diff is an
insufficient collaboration artifact. Sun et~al. show that code change rationale
is fragmented across commits, issues, and PRs, and that no single artifact
captures all rationale components~\cite{sun2026rationale}. Pirouzkhah et~al.
analyze 80k PRs and find that description elements have different roles:
purpose and code explanations preserve rationale and history, while desired
feedback type predicts reviewer engagement and
acceptance~\cite{pirouzkhah2026prdescription}. KPR generalizes this line of
work to agent-mediated development by treating rationale, evidence, rejected
alternatives, and human corrections as first-class contents of the contribution
artifact rather than as optional prose around a diff.

\subsection{Agents as Participants in Software Work}

Software engineering bots have long mediated collaborative development. Wessel
et~al. characterize bots in popular projects hosted on code collaboration
platforms and show that they automate predefined tasks that assist contributors
and integrators~\cite{wessel2018powerbots}. Santhanam et~al.'s systematic
mapping study shows that bots span many software engineering activities,
confirming that automation is already embedded in development
workflows~\cite{santhanam2022botsmapping}. KPR extends this tradition from
task automation to workflow mediation: different agents extract knowledge,
transform it into reviewer-facing forms, and regenerate implementation inside
the receiving project.

LLM programming-assistant studies explain why local agent sessions can be more
informative than the final patch. Barke et~al. find that programmers use
code-generating models in both acceleration and exploration
modes~\cite{barke2023groundedcopilot}. Ross et~al. show that conversational
LLM interaction can support extended, code-grounded, multi-turn development
work rather than only single-shot generation~\cite{ross2023programmersassistant}.
SWE-chat provides a complementary trace-level view: in roughly 6,000 real
coding-agent sessions, Baumann et~al. report that only 44\% of agent-produced
code survives into user commits and that users push back against agent output
in 44\% of turns~\cite{baumann2026swechat}. This evidence supports a core KPR
assumption: the useful contribution is not only the final generated code, but
also the corrections, validation attempts, and abandoned alternatives produced
during local exploration.

\subsection{Agent-authored PRs and Input Quality}

Recent empirical studies show that agentic PRs are already a concrete
collaboration phenomenon. Njoku et~al. study more than 40k PRs and compare
agent-authored and human-authored contributions across integration outcomes,
structural characteristics, and collaboration
signals~\cite{njoku2026codeauthors}. Ehsani et~al. focus on failed agentic PRs
and report that success varies substantially by task
type~\cite{ehsani2026where}. Peralta et~al. show that rejection is not always
equivalent to agent failure: rejected cases may reflect workflow mismatch or
missing observable rationale as well as incorrect implementation
behavior~\cite{peralta2026why}.

Issue-quality work points to the same dependency on upstream context. Sayagh
finds that issues leading to merged Copilot PRs tend to be shorter,
well-scoped, and explicit about relevant artifacts and implementation guidance,
while issues with significant external context are associated with lower merge
rates~\cite{sayagh2025copilotissue}. KPR responds to this input-quality
problem by making the contribution package richer than a sparse issue but more
controlled than an external patch. It asks agents to transform local
exploration into a reviewable knowledge package before the project invests in
implementation review.

\subsection{Spec-Driven and Issue-to-Code Workflows}

Issue-to-code and spec-to-code workflows ask whether agents can transform
natural language requirements into implementation. Spec-driven development
pushes this further by treating specifications, rather than code, as the
primary artifact~\cite{piskala2026specdriven}. Spec Kit Agents adds
repository-grounded discovery and validation hooks to spec-driven agentic
workflows and reports improved judged quality across a controlled task
set~\cite{taghavi2026speckitagents}. Practice-oriented workflows also move in
this direction: Warp/Oz for OSS uses staged labels and review gates to move
contributions from issue discussion toward specification and
implementation~\cite{warpOzContributing}.

These systems are close to KPR because they separate structured intent from
generated code. The difference is the trust boundary and the source of the
specification. Spec-driven development usually assumes an internal or otherwise
trusted specification process. KPR begins from external local exploration that
may include generated code, tests, failed attempts, private context, and
possibly contaminated traces. The receiving project must therefore decide not
only whether an agent can implement a specification, but whether agents can
convert external exploration into project-native knowledge without granting
external implementation authority.

\subsection{Provenance, Safety, and Project Context}

KPR depends on provenance because the receiving project must understand which
claims came from which materials. Wang et~al. argue that final-answer accuracy
is insufficient for agent trust: evidence, tool outputs, memory, observations,
intermediate claims, and final actions must be connected throughout
execution~\cite{wang2026traces}. KPR applies this principle to software
collaboration by requiring knowledge-package claims to retain selective
provenance without forcing reviewers to read entire raw traces.

Security and context-management work also constrain the design. Prompt
injection research shows that agentic coding assistants have broad attack
surfaces across tools, skills, and protocols~\cite{maloyan2026promptinjection}.
Repository context files illustrate that simply adding more context is not
enough: Gloaguen et~al. find that \texttt{AGENTS.md}-style files can reduce
task success and raise inference costs even when agents follow their
instructions~\cite{gloaguen2026agentsmd}. KPR therefore cannot be a workflow
that sends all external context directly to a project agent. It must separate
validated evidence, untrusted trace material, human-confirmed claims, and
project-side policy constraints.

\subsection{Positioning KPR}

KPR does not claim that knowledge extraction, specification writing,
provenance tracking, or agentic code generation are individually new
capabilities. Its claim is a workflow-level recombination for software
collaboration across trust boundaries. External code is treated as evidence
rather than as the default merge candidate; a human-confirmed knowledge package
is the artifact that crosses the boundary; a project-owned inner trusted coding
agent regenerates candidate implementation under project policy; and
project-side humans separately gate knowledge acceptance and code integration.

The strongest equivalence critique is that KPR is spec-driven development for
untrusted contributors. KPR intentionally borrows the idea that structured
intent can guide agentic implementation, but it adds a governance layer around
source, evidence, and authority. Its central question is not simply whether an
agent can implement a spec. It is whether a receiving project can use agents to
extract, audit, translate, and act on external development knowledge while
keeping implementation authority, policy enforcement, and merge decisions
inside the project boundary.

\section{The Problem: Collaboration Burden, Not Code Scarcity}

The traditional PR asks project-side reviewers to review code first. This is
efficient when the proposed change is small, well-scoped, and authored within
familiar project norms. It becomes less efficient when an external coding agent
produces a large, plausible patch whose underlying intent is unclear. In such
cases, the expensive work is not typing code. It is reconstructing the problem,
understanding the external collaborator's exploration, translating it into
project language, checking policy fit, and deciding whether the project should
own the change.

\subsection{Code Is a Lossy Record of Intent}

Modern agent-assisted development is rarely a one-step process. A developer may
begin with an incomplete request, let a local agent generate a first attempt,
notice that the agent misunderstood the boundary, add constraints, run tests,
reject a design, and eventually converge on a change that works for their local
need. The final diff preserves one outcome of that process. It usually does not
preserve why the external collaborator changed their mind, which alternatives
failed, which tests represented the user need, or which assumptions remained
uncertain.

This loss matters because project-side reviewers review more than syntax. They
review fit. They ask whether the request belongs in the project, whether it
violates past design decisions, whether it creates future maintenance cost, and
whether the evidence is strong enough to justify the change. When the PR exposes
only the final code, project-side reviewers must infer those answers by
reverse-engineering the patch. That inference work is precisely the kind of
information processing that agents can assist with, if the workflow provides
the right intermediate artifact.

\subsection{Raw Agent Traces Are Not the Answer}

One might respond by attaching the full local agent conversation to the PR.
That is insufficient. Raw traces can contain stale conclusions, irrelevant
trial-and-error, copied documents, secrets, private environment details, and
untrusted instructions. They can also be too long for human review. Asking
project-side reviewers to read dozens of chat turns and tool outputs would move
the bandwidth problem from code review to trace review.

The missing artifact is therefore not raw trace and not raw code. It is a
reviewable and transformable knowledge object: compact enough for project-side
reviewers, structured enough for agents, explicit about uncertainty, and backed
by provenance when a claim needs inspection. The same KPR package should be
able to support multiple agent-generated views: a short reviewer brief, a
design memo, a test plan, a risk checklist, or an implementation prompt for the
inner trusted agent.

\subsection{The Trust Boundary}

KPR targets software collaboration that crosses a trust boundary. External
collaborators may be well-intentioned, but the receiving project cannot assume
that their local agent, environment, dependencies, or generated patch satisfy
project policy. In a traditional PR, external code crosses the boundary and
becomes the merge candidate. In an issue-based flow, sparse knowledge crosses
the boundary but often lacks implementation evidence. KPR proposes a middle
path: let knowledge cross the boundary, but regenerate mergeable code inside
the project.

\subsection{Burden Redistribution and Cost Accounting}

KPR should not be understood as a free reduction in work. It adds stages:
knowledge extraction, view generation, human confirmation, knowledge review,
project-side regeneration, fidelity checking, and ordinary code review. The
claim is therefore conditional. KPR is useful only when agents can absorb
routine, auditable transformation work and when that absorbed work is cheaper
than asking project-side developers to reverse-engineer external code, repair
architecture mismatches, and rediscover constraints late in review.

Table~\ref{tab:cost-accounting} makes this cost model explicit. For small
mechanical fixes, the KPR path is likely too heavy. For high-context changes
that cross a trust boundary, the added agentic stages may be justified if they
turn unstructured external exploration into project-native review material and
reduce later rework. This is a hypothesis to evaluate, not a property of the
workflow by definition.

\begin{table*}[t]
  \caption{A cost-accounting view of KPR relative to ordinary code-first
  review. KPR adds stages; its value depends on whether agent-handled work
  replaces more expensive project-side reverse engineering and rework.}
  \label{tab:cost-accounting}
  \centering
  \begin{tabular}{p{0.18\textwidth}p{0.24\textwidth}p{0.24\textwidth}p{0.24\textwidth}}
    \toprule
    Work item & Traditional PR & KPR & Cost that must be measured \\
    \midrule
    Intent and rationale & Often inferred from diff, PR text, issue links,
    and discussion. & Extracted from local trace, external diff, tests, and
    human corrections into a knowledge package. & Collaborator confirmation
    time; agent extraction cost; reviewer trust in extracted claims. \\
    First project-side judgment & Reviewers inspect external code and intent
    together. & Reviewers first judge problem fit, evidence, constraints, and
    responsibility. & Time to reject, approve, or ask for clarification. \\
    Implementation & External code is the candidate patch, possibly revised by
    reviewers or maintainers. & A project-owned inner trusted coding agent
    regenerates candidate code under project policy. & Agent runtime,
    regeneration failures, architecture fit, and security-policy compliance. \\
    Fidelity checking & Reviewers compare the patch to intended behavior,
    often reconstructing intent during review. & Reviewers compare
    project-generated code to the accepted knowledge package and acceptance
    criteria. & Fidelity errors, missing constraints, and extra review surface. \\
    Net value condition & Efficient when the external patch is small,
    trustworthy, and project-native. & Potentially useful when external code is
    high-context, cross-module, policy-sensitive, or architecture-mismatched. &
    End-to-end human time, clarification rounds, rework, defects, and perceived
    governance control. \\
    \bottomrule
  \end{tabular}
\end{table*}

\section{Knowledge-Based Pull Requests}

Figure~\ref{fig:kpr-overview} summarizes the difference between issue-based,
traditional PR, and KPR workflows. KPR has seven stages:

\begin{enumerate}
  \item An external collaborator works locally with a coding agent to explore,
  implement, and test a change.
  \item An extraction agent cleans the local trace, local diff, tests, logs,
  and human corrections, then distills them into a KPR package.
  \item A transformation agent renders the package into reviewer-facing forms:
  a concise reviewer brief, design memo, risk checklist, test plan, or
  implementation brief, depending on project preference.
  \item The collaborator reviews and confirms the package and its generated
  views, removing private or irrelevant material and taking responsibility for
  the claims they submit.
  \item Project-side reviewers perform a knowledge review: they may reject the
  package, ask for clarification, or approve it for project-side implementation.
  \item A project-owned inner trusted coding agent regenerates candidate code
  under repository context, tests, conventions, and security policy.
  \item Project-side reviewers perform ordinary code review, CI, security review,
  and merge/reject decisions on project-generated code.
\end{enumerate}

This design separates two decisions that traditional PRs often collapse:
\emph{Should this knowledge enter the project?} and \emph{Is this code an
acceptable implementation?} The first decision concerns product fit, user
need, constraints, evidence, and responsibility. The second concerns code
quality, tests, maintainability, and integration. Agents carry the routine
translation, summarization, and regeneration work around those decisions. The
decisions themselves remain human-governed.

\subsection{Agent Roles}

KPR is agent-centered in the sense that it assigns different agent roles to
different kinds of collaboration work:

\begin{description}
  \item[Local exploration agent:] helps the external collaborator discover the
  need, try implementations, run tests, and identify edge cases.
  \item[Extraction agent:] turns messy local traces, diffs, and test evidence
  into a structured knowledge package with provenance.
  \item[Transformation agent:] converts the package into the forms project-side
  reviewers prefer to review, such as a design memo, test plan, risk checklist,
  or short decision brief.
  \item[Inner trusted coding agent:] operates inside the project boundary and
  regenerates mergeable code under repository context, conventions, and security
  policy.
\end{description}

These roles redistribute collaboration burden by letting agents handle
high-volume, auditable information work while project-side humans retain
control over acceptance and merge. The redistribution is valuable only when the
agent-handled work can be checked and when it replaces more expensive human
coordination or rework.

\subsection{What KPR Is Not}

KPR is not a replacement for code review, and it is not a demand that project
reviewers accept an external collaborator's original implementation. The
opposite is true: KPR lets the receiving project accept, reject, or modify the
underlying idea without accepting the external patch as authoritative. It does
not claim that a design doc is sufficient for merge. It does not claim that
project-side regeneration eliminates license or security obligations. It also
does not require every PR to become a knowledge package. Small bug fixes,
compatibility patches, and low-risk mechanical changes may still be more
efficiently handled as ordinary code PRs.

KPR is intended for changes where the cost of reconstructing intent from code
is high: cross-module features, behavior changes discovered through local use,
high-context bug fixes, security-sensitive contributions, and external patches
whose implementation may not match project architecture even if the underlying
need is valid.

\subsection{Why Call It a Pull Request?}

The name is intentionally imperfect. In KPR, the artifact crossing the trust
boundary is not a request to pull external code directly into the codebase. It
is a request to pull a body of validated knowledge into the receiving project's
governance process. The term ``pull request'' is retained because KPR preserves
the social function of a PR: an external party asks a project to consider an
integration-relevant contribution, project-side reviewers decide whether it
fits, and the accepted result may eventually enter the shared system.

For organizations that do not use PR terminology, the same mechanism could be
called a knowledge contribution, evidence package, or collaboration gateway
submission. The important property is not the label. It is the separation
between knowledge acceptance and implementation acceptance across a trust
boundary.

\subsection{Implementation Authority}

The central governance claim of KPR is that implementation authority should
remain with the project when the code will become part of the shared codebase.
External code can be valuable evidence. It can show that a problem is real, that
a behavior is feasible, and that certain edge cases matter. But in KPR, that
code is not the default artifact to merge. It is a knowledge source for the
project's inner trusted coding agent.

This matters because the inner trusted coding agent can operate with the
project's tests, style rules, architecture, dependency policy, security rules,
and reviewer constraints. It can also be instrumented, sandboxed, and audited
by the project. KPR does not make the inner trusted agent infallible. It makes
control over implementation explicit.

In this sense, KPR changes the role of the original PR-like work. The external
implementation is useful because it helps agents understand what the external collaborator
was trying to accomplish. It is not useful because the project must trust it.
The project can use agents to understand the idea, translate it into internal
terms, and then ask an inner trusted coding agent to re-develop the change from
the accepted knowledge under project rules. Every stage that affects mergeable
code remains inside the project's own workflow.

\subsection{Regeneration Fidelity and Cost}

Project-side regeneration is the most fragile part of KPR. An inner trusted
coding agent may fail to reconstruct a faithful implementation from the accepted
knowledge package, especially for cross-module behavior changes. More project
context is not automatically better: repository-level instruction files can
raise inference cost and reduce task success in some settings~\cite{gloaguen2026agentsmd}.
KPR therefore treats regeneration as an empirical question rather than as a
guarantee.

Fidelity should be defined against the accepted knowledge package, acceptance
criteria, tests, and reviewer-approved constraints, not against the external
diff. In the strictest evaluation mode, the inner trusted agent should not see
the external implementation; this tests whether the package is sufficient as a
knowledge transfer artifact. In a more practical mode, the external diff may be
available as non-authoritative evidence, while the project still requires the
agent to produce code under internal policy. Both modes are useful: the first
tests sufficiency, and the second tests whether KPR improves real collaboration
when projects allow reference implementations.

\subsection{Contributor Credit and Authorship}

KPR changes the contributor bargain. The external collaborator does additional
knowledge work and may not become the author of the final merged code if the
project regenerates the implementation. Without explicit credit, this can turn
KPR into a worse deal for contributors than an ordinary PR.

A KPR workflow should therefore record contribution credit separately from code
authorship. The generated implementation can cite the KPR package in its
metadata, discussion, or commit message, and projects can distinguish
``knowledge package by'' from ``implementation generated by'' and ``reviewed
by'' records. DCO, CLA, and internal approval processes should likewise
distinguish the provenance of submitted knowledge from the authorship of
project-generated code. This credit model does not solve licensing by itself,
but it prevents regeneration from erasing the external collaborator's role.

\section{KPR Artifact Schema}

A KPR package should be structured enough that agents can transform it and
project-side reviewers can skim it, but not so rigid that every project must
adopt the same template. Table~\ref{tab:kpr-schema} proposes a starting
schema. A project can choose which reviewer-facing views it wants agents to
generate from the same package: a short decision brief, a design memo, a test
plan, a security checklist, or an implementation plan for the inner trusted
agent.

\begin{table*}[t]
  \caption{Candidate KPR package schema.}
  \label{tab:kpr-schema}
  \centering
  \begin{tabular}{p{0.18\textwidth}p{0.48\textwidth}p{0.25\textwidth}}
    \toprule
    Field & Purpose & Example review question \\
    \midrule
    Summary & Briefly states the intended change and why it matters. &
    What problem is this package asking the project to solve? \\
    Motivation and scenarios & Describes concrete user or system scenarios
    where the need appears. & Is this a real project need or a local-only
    preference? \\
    Expected behavior & Defines observable behavior after successful
    implementation. & How would reviewers know the change works? \\
    Constraints and non-goals & States compatibility, performance, security,
    API, UX, and scope boundaries. & What must not be broken or expanded? \\
    Evidence & Links reproduction steps, local tests, logs, screenshots,
    benchmarks, or user observations. & What has actually been validated? \\
    External implementation notes & Summarizes the local diff as evidence,
    not as mergeable code. & What did the external implementation teach us? \\
    Reviewer-facing views & Specifies generated forms such as design memo,
    risk checklist, test plan, or implementation brief. & Is the package shown
    in the form this project prefers? \\
    Alternatives considered & Records rejected approaches and why they were
    rejected. & Are reviewers repeating already-failed exploration? \\
    Provenance map & Connects claims to cleaned trace snippets, files, tests,
    or human edits. & Where did each important claim come from? \\
    Safety and license notes & Identifies secrets removed, copied material,
    prompt-injection risk, third-party code, and license concerns. & Is there
    untrusted or legally sensitive input? \\
    Human confirmation & States what the collaborator reviewed and accepts
    responsibility for. & Is there a human accountable for the package? \\
    Open questions & Lists uncertainties requiring project-side judgment. & What
    needs project-side clarification before implementation? \\
    \bottomrule
  \end{tabular}
\end{table*}

The table is a candidate superset, not a mandatory form. A minimal KPR should
contain summary, motivation and scenarios, expected behavior, constraints and
non-goals, evidence, external implementation notes, safety and license notes,
human confirmation, and open questions. Reviewer-facing views, detailed
alternatives, and fine-grained provenance can be generated or requested when
the project needs them. This layered design matters because an oversized KPR
template could recreate the same burden it is meant to reduce.

This schema intentionally separates evidence from implementation. A local diff
is included only as an implementation note and knowledge source. The KPR should
not say ``merge this patch.'' It should say ``this local exploration suggests
the following need, constraints, and evidence; please decide whether the
knowledge should enter the project and, if so, regenerate the implementation
inside the project boundary.''

\subsection{Human Confirmation}

Human confirmation is required because an automatically summarized trace can
distort intent. A local agent may hallucinate a requirement, omit a failed
alternative, or overstate test coverage. The collaborator must check the package
before submission. This is not merely a UI step; it is the responsibility
handoff. The project should treat an unconfirmed KPR as machine output, not as a
trusted collaboration artifact.

\subsection{Provenance Without Trace Dumping}

The provenance map should support selective inspection. Project-side reviewers
should not need to read the entire local transcript. But when a package claims
``this edge case was validated'' or ``this approach was rejected for
performance reasons,'' the package should point to the relevant cleaned trace,
test output, or local diff segment. This design follows the provenance
principle that trust depends on process-level evidence, not only final
outputs~\cite{wang2026traces}.

\section{Prototype Architecture: A Collaboration Gateway}

KPR can be prototyped without rebuilding a full code collaboration platform.
The minimal prototype is a collaboration gateway that sits between local
agent-assisted development and project-side regeneration. Its purpose is not to
create a better AI summary.
Its purpose is to move structured, auditable transformation work away from
project-side reviewers when an agent can do that work: extracting knowledge,
presenting it in preferred forms, checking it against policy, and preparing an
implementation brief for the inner trusted coding agent. It does not remove
reviewer judgment.

\subsection{Inputs}

The gateway ingests:

\begin{itemize}
  \item a local coding-agent transcript or structured session log;
  \item the local diff produced during exploration;
  \item tests, commands, logs, screenshots, or benchmark outputs;
  \item project metadata such as issue links, target branch, and repository
  contribution or collaboration rules;
  \item human notes supplied by the external collaborator.
\end{itemize}

These inputs are untrusted by default. The gateway should not execute arbitrary
external code in the project environment. It should treat the local diff as a
document to analyze, not as a patch to apply.

\subsection{KPR Generation}

The gateway then produces a draft KPR package and a set of reviewer-facing
views. This requires several agentic subsystems:

\begin{itemize}
  \item \textbf{Sanitization and tainting:} remove secrets, private paths,
  irrelevant logs, and obvious prompt-injection content, while retaining taint
  labels that mark all external trace material as untrusted.
  \item \textbf{Trace summarization:} identify goals, constraints, validation
  steps, rejected alternatives, and unresolved questions.
  \item \textbf{Diff abstraction:} summarize what the local patch did without
  treating it as authoritative implementation.
  \item \textbf{Evidence linking:} connect package claims to tests, commands,
  trace snippets, and diff regions.
  \item \textbf{Reviewer-view generation:} render the same package as a short
  decision brief, design memo, risk checklist, test plan, or implementation
  brief according to project preference.
  \item \textbf{Policy precheck:} compare structured package claims with
  visible project collaboration rules, security constraints, and known
  architectural boundaries before asking project-side reviewers to spend
  attention.
  \item \textbf{Human editing:} require the collaborator to confirm, correct, or
  delete generated claims.
\end{itemize}

The output is a signed or otherwise attributable KPR package plus
reviewer-facing views. In a code collaboration platform workflow, this
package could be attached to an issue, submitted through a new contribution
type, or placed in a project-specific queue. The project-side reviewer should
be able to start from the view they prefer, while retaining access to the
underlying package and provenance links.

\subsection{Project-Side Regeneration}

If project-side reviewers approve the knowledge gate, the project-owned inner trusted
coding agent receives the KPR package plus repository context. The agent should
operate in a clean checkout, under project-configured tools and policy. It can
consult the external local diff only as a reference artifact if the project
allows it. A stricter evaluation mode can hide the external diff from the inner
agent to test whether the KPR package alone is sufficient.

By default, the inner trusted agent should receive structured claims, evidence
links, and reviewer-approved constraints, not the raw external trace. Raw and
cleaned traces remain external data for selective inspection. This does not
eliminate prompt-injection risk; it narrows the pathway by which untrusted
language can influence project-controlled execution.

This step is where project control becomes operational. The inner trusted
coding agent should not simply copy external code into the repository. It
should treat the accepted KPR as a requirements-and-evidence bundle, then
re-develop the change under project-owned tests, internal conventions,
dependency policy, security checks, and reviewer instructions.

The inner trusted coding agent produces:

\begin{itemize}
  \item a candidate patch;
  \item a mapping from KPR claims to implemented files/tests;
  \item unresolved questions or missing constraints;
  \item CI, test, and static-analysis results;
  \item a provenance report describing which inputs influenced the patch.
\end{itemize}

Project-side reviewers then perform ordinary review. KPR changes what comes
first; it does not remove the final gate. The intended value is that final
review starts from project-generated code whose rationale, evidence, and policy
assumptions were already organized by agents.

\section{Controlled Simulation Pilot}

To check whether KPR can be operationalized as a study artifact, we built a
minimal controlled simulation pilot over seven merged public PRs. This pilot is
not an end-to-end validation of KPR and does not involve project maintainers.
It tests a smaller claim: whether real PR material can be converted into KPR
packages and evaluated under controlled information loss and synthetic
poisoning conditions.

\subsection{Pilot Corpus}

The pilot corpus contains seven merged public PRs selected to be small enough
for manual inspection while still containing real PR structure. Each PR has a
public body, file-level patch data, at most five changed files, and at most 350
added plus deleted lines. The sample includes small API exposure, test
regression, documentation/testing, automated maintenance, and workflow-security
changes. This sample is intentionally modest; it is used to exercise the KPR
artifact and scoring protocol, not to estimate population-level effects.

\subsection{Study Conditions}

For each PR, we generated five artifact conditions:

\begin{description}
  \item[Normal summary:] a baseline artifact containing the visible PR title,
  body summary, and changed-file list.
  \item[KPR package:] a structured package containing summary, motivation,
  expected behavior, constraints, evidence, external implementation notes,
  provenance, safety notes, confirmation status, and open questions.
  \item[Description ablation:] the PR body is redacted before KPR generation;
  the package must rely on title, file names, and compact patch signals.
  \item[Diff ablation:] patch content is redacted before KPR generation; the
  package must rely on title, body, file names, and change statistics.
  \item[Synthetic poisoned patch:] a non-executable synthetic malicious hunk is
  inserted into the external patch text to test whether the KPR artifact can
  represent it as untrusted, non-legitimate external code.
\end{description}

We scored the generated artifacts with a conservative author rubric from 0 to
2 on intent correctness, constraint correctness, evidence traceability,
implementation sufficiency, and poison rejection for the poisoned condition.
A score of 2 indicates that the property is clearly present and supported by
available evidence. Because the scoring is author-scored and the sample is
small, the results should be read as an instrumentation check rather than as a
claim about maintainer productivity.

\begin{table}[t]
  \caption{Author-scored results from the seven-PR controlled simulation
  pilot. Cells report artifacts scored 2 out of all rated artifacts in that
  condition.}
  \label{tab:pilot-results}
  \centering
  \begin{tabular}{lrrrr}
    \toprule
    Condition & Intent & Evidence & Impl. & Poison \\
    \midrule
    Normal summary & 7/7 & 0/7 & 0/7 & -- \\
    KPR package & 7/7 & 7/7 & 6/7 & -- \\
    Description ablation & 7/7 & 7/7 & 6/7 & -- \\
    Diff ablation & 7/7 & 0/7 & 5/7 & -- \\
    Synthetic poisoned KPR & 7/7 & 7/7 & 6/7 & 7/7 \\
    \bottomrule
  \end{tabular}
\end{table}

\subsection{Pilot Findings}

The pilot supports three limited observations. First, ordinary summaries and
KPR packages both preserved visible intent in this small sample; the easy
intent scores are partly explained by descriptive PR titles and small changes.
Second, KPR packages made evidence traceability explicit when patch signals
were available, while normal summaries did not separate claims, evidence,
provenance, and safety. Third, compact external-code signals were useful:
description-ablation KPR packages still preserved enough implementation
knowledge for six of seven PRs, while diff-ablation packages were weaker and
only reached implementation sufficiency in five of seven cases. The automated
plugin-list update was the clearest failure mode because a compact summary
does not contain enough information to regenerate an exact list update.

The synthetic poisoning condition should be interpreted narrowly. All seven
poisoned KPR packages rejected the synthetic hunk because the harness marks it
as non-legitimate external code. This shows that the KPR artifact can represent
untrusted external-code distinctions and can be stress-tested, not that KPR is
secure against real adversarial prompt injection or malicious code.

The pilot also exposes a weakness in the current schema: constraint specificity
remained low across conditions. The packages captured generic constraints such
as changed-file scope and non-authoritative external code, but they did not
extract rich project-specific non-goals. This result reinforces the need for
maintainer studies and project-specific policy integration rather than
supporting a strong validation claim.

\section{Evaluation Agenda}

The strongest future version of this work is empirical. The first step should
be a minimal gateway prototype and a formative study with a small number of
experienced reviewers or maintainers. The goal of that first study is not to
prove that KPR wins. It is to learn whether the artifact is understandable,
whether reviewers trust the provenance model, where regeneration fails, and
whether the workflow creates unacceptable contributor or reviewer burden.

\subsection{Prototype Milestones}

A minimal prototype should implement only the pieces needed to test knowledge
transfer:

\begin{itemize}
  \item ingest a local agent transcript, external diff, tests, and human notes;
  \item generate a layered KPR package with required fields and optional views;
  \item preserve provenance links from package claims to supporting material;
  \item require collaborator confirmation before submission;
  \item run an inner trusted coding agent in a clean project checkout;
  \item report where generated code satisfies, misses, or contradicts accepted
  KPR claims.
\end{itemize}

\subsection{Conditions and Ablations}

KPR should be compared with strong baselines, not only with sparse issues or
low-quality PR descriptions. A useful factorial design would vary both the
knowledge artifact and the regeneration policy:

\begin{description}
  \item[Structured PR or issue:] a strong template containing rationale,
  reproduction steps, acceptance criteria, tests, and risk notes.
  \item[KPR without regeneration:] reviewers receive the KPR package and
  generated views, but the external diff remains the candidate patch.
  \item[KPR with reference regeneration:] the inner trusted agent regenerates
  code and may inspect the external diff as non-authoritative evidence.
  \item[KPR with blind regeneration:] the inner trusted agent receives only the
  accepted KPR package and project context, not the external diff.
\end{description}

This design separates the value of structured knowledge from the value and cost
of project-side regeneration. If KPR without regeneration performs as well as
full KPR, then regeneration may be unnecessary. If blind regeneration fails but
reference regeneration works, then external code may be necessary as evidence
even when it should not be the merge candidate.

\subsection{Tasks}

Tasks should include multiple risk profiles:

\begin{itemize}
  \item documentation or examples, where intent is usually easy to validate;
  \item small bug fixes, where reproduction evidence matters;
  \item behavior-changing features, where product fit and boundary conditions
  dominate;
  \item cross-module refactors, where architecture and tests matter;
  \item security- or privacy-sensitive changes, if qualified reviewers and
  safeguards are available.
\end{itemize}

The study should use real repositories or realistic forks with meaningful
policy and access boundaries. Enterprise-style simulations are useful only if
they preserve the trust-boundary conditions that motivate KPR; otherwise, they
should be presented as speculative extensions rather than evidence for
organizational deployment.

\subsection{Measures}

Quantitative measures should include:

\begin{itemize}
  \item project-side reviewer time to first judgment;
  \item number of clarification rounds;
  \item collaborator time spent preparing, correcting, and confirming the KPR;
  \item agent runtime, token use, and failed regeneration attempts;
  \item reviewer confidence in the problem statement, constraints, and
  non-goals;
  \item usefulness of generated reviewer-facing views;
  \item implementation fidelity relative to the accepted knowledge package,
  acceptance tests, and reviewer-approved constraints;
  \item CI and test pass rates;
  \item number of security, license, or policy issues discovered;
  \item rework before merge;
  \item post-merge defects or reversions where observable.
\end{itemize}

Qualitative measures should ask reviewers what information helped, what was
missing, whether the agent-generated views matched their preferred working
style, when the package felt trustworthy, whether regeneration made review
easier or harder, and whether contributor credit felt adequate. KPR should be
considered successful only if it improves project-side judgment enough to
justify the cost of preparing, reviewing, and regenerating from the package.

\subsection{Key Hypotheses}

The evaluation should not assume KPR wins. It should test specific hypotheses:

\begin{itemize}
  \item KPR reduces time to reject changes whose underlying need does not fit
  project goals.
  \item KPR reduces clarification rounds for high-context behavior changes.
  \item KPR improves reviewer confidence in intent and constraints relative to
  strong structured PR or issue templates.
  \item Agent-generated reviewer-facing views reduce the effort required to
  understand local exploration.
  \item Project-side regeneration improves architecture and policy alignment
  for changes whose external implementation does not fit project norms.
  \item KPR adds overhead for low-risk tasks where ordinary PRs are already
  efficient.
\end{itemize}

Negative results would be useful. If structured templates perform as well as
KPR, the trace and local diff may not justify their cost. If inner trusted
agents cannot regenerate faithful patches from KPR packages, the package schema
is insufficient. If project-side reviewers find KPR harder to review than code,
or if contributors reject the credit model, the workflow fails its central
purpose.

\section{Risks and Limitations}

\subsection{Spec Spam}

KPR may move spam from code to knowledge. A malicious or careless collaborator
could generate plausible packages with weak evidence. KPR should therefore
require evidence, provenance, and human accountability. Projects should reject
packages without reproduction steps, acceptance criteria, or clear
responsibility.

\subsection{False Confidence}

A structured package can look more credible than it is. This is especially
dangerous if a summarizer omits uncertainty or converts speculation into
assertion. KPR tools should preserve uncertainty and make unsupported claims
visible. Reviewers should be able to ask the package: which claims are human
confirmed, which are inferred from trace, and which are unsupported?

\subsection{Prompt Injection and Trace Contamination}

External traces can contain adversarial instructions. A KPR gateway must not
feed raw traces directly into a project-owned agent. It should sanitize,
segment, and label trace material, and the inner trusted agent should be
instructed to treat external artifacts as untrusted evidence rather than
operational instructions. Sanitization is not a solved problem: injected
instructions may survive summarization, paraphrase, or evidence extraction.
KPR therefore needs defense in depth, including taint labels, allowlisted
claim extraction, human review of high-impact claims, isolated regeneration
environments, and audit logs that show which external materials influenced the
inner trusted agent. This is an architectural requirement, not a prompt-writing
detail.

\subsection{License and Authorship}

Regenerating code inside the project does not automatically remove license or
authorship concerns. If the KPR package contains copied code, proprietary
snippets, or generated output derived from incompatible sources, those
materials can still influence the final patch. KPR systems need provenance and
possibly license checks for upstream materials, not only for the final code.
They also need contributor-credit policies so that regeneration does not erase
the external collaborator's intellectual and evidentiary contribution.

\subsection{Contributor Burden}

KPR can ask more of contributors than an ordinary PR. A collaborator may need
to review generated summaries, remove private content, confirm evidence,
answer open questions, and accept that their local implementation will not be
merged directly. Projects should reserve KPR for cases where this additional
work is justified and should make the credit model visible before asking for a
knowledge package.

\subsection{Reviewer Burden}

KPR could add another artifact for project-side reviewers to read. Its value
depends on whether that artifact reduces more expensive work later. The
evaluation must measure end-to-end burden for both contributors and
project-side reviewers, not merely whether KPR packages look well structured.

\section{Discussion}

KPR reframes software collaboration across trust boundaries. Instead of asking
only ``Can an agent generate a patch from a request?'', it asks ``Which parts
of cross-boundary collaboration can agents handle, at what cost, while the
receiving project keeps control?''

Issue-based workflows send sparse knowledge and no code. Traditional PRs send
external code and ask project-side reviewers to judge both intent and
implementation at once. KPR sends a knowledge package whose sources may include
external code, but reserves mergeable code generation for the project side.
This creates a different socio-technical contract. External collaborators
contribute validated knowledge. Agents summarize, translate, check, and
regenerate. Project-side humans govern product fit, responsibility, policy, and
final acceptance.

The workflow also changes what ``good collaboration'' means. In a code-first
PR, the external collaborator is rewarded for producing a patch that passes
tests and looks mergeable. In KPR, the collaborator is rewarded for making the
problem, boundary, evidence, and responsibility legible enough that agents can
transform it into project-native forms. This may be especially valuable when
the external collaborator has real user context but lacks deep knowledge of the
project's internal architecture. It is also a reason KPR must make credit
explicit; otherwise, it asks contributors to do knowledge work while hiding
their role behind project-generated code.

KPR also creates a tension with the knowledge-transfer value of code review.
Modern review is not only defect detection; it helps teams develop shared
understanding and alternative designs. Delegating information work to agents
should not remove the moments where project-side reviewers learn why a change
matters. KPR therefore should automate transformation, summarization, and
cross-checking, but preserve human attention for product fit, architecture,
responsibility, credit, and final integration decisions.

KPR also suggests a broader research direction: collaboration artifacts should be
designed for human-agent collaboration, not merely for human review or agent
execution alone. A KPR package must be readable by project-side reviewers,
transformable by agents, auditable through provenance, and accountable through
human confirmation. The design challenge is not to replace project-side
developers. It is to let them spend attention only where their judgment is
actually needed, while measuring whether the agent-mediated workflow truly
reduces, or merely relocates, collaboration cost.

\section{Conclusion}

As coding agents make code cheaper to produce, software projects may become
limited less by implementation capacity than by project-side attention, trust,
and governance. This is visible in open source, but it also matters in
enterprise, vendor, contractor, and customer-driven collaboration. The
traditional pull request exposes reviewers first to external code, even when
the most valuable artifact is the knowledge produced during local exploration.
KPR proposes a different boundary artifact: a
human-confirmed, provenance-bearing knowledge package that agents can
summarize, translate, check, and use for project-side implementation.

KPR is not a finished solution. It is a product and research hypothesis. Its
value depends on whether agents can convert messy local exploration into
reviewable project-side knowledge, whether project-owned inner trusted agents
can regenerate faithful implementations under repository policy, and whether
the total burden is lower than strong structured PR or issue workflows. This
paper defines the workflow, artifact schema, cost model, prototype architecture,
minimal controlled simulation pilot, and evaluation agenda needed to test that
hypothesis. The pilot shows that KPR packages can be instantiated on real PR
material and stress-tested under description ablation, diff ablation, and
synthetic poisoning, but it does not validate maintainer burden reduction or
project-side regeneration. If validated by stronger studies, KPR could become
one way for software collaboration governance to keep pace with
agent-accelerated code production: project-side humans review ideas, evidence,
policy fit, credit, and final project-generated code, while agents handle the
summarization, translation, and re-development work in between.

\appendix

\section{Ethical Statement}

This paper is a conceptual research framework and evaluation agenda. It does
not report human-subjects research, collect or analyze private user data,
deploy a system to users, or conduct experiments involving participants. The
minimal controlled simulation pilot uses only public PR metadata and patches
from public repositories. Therefore, this work does not raise direct ethical
concerns and did not require additional ethics review. Future empirical
evaluations of KPR, especially studies involving developers, organizations,
proprietary repositories, or real collaboration traces, should be reviewed
separately under the applicable institutional and organizational ethics
processes.

\section{LLM Usage Statement}

Large language models were used as writing assistance to improve grammar,
clarity, and organization. They were also used to help search for and identify
related work, draft study materials, and assist with analysis-script
development. The controlled simulation pilot scores reported in this paper are
author-scored feasibility scores, not an independent LLM evaluation. The author
selected the final framing, reviewed and revised the text, checked the cited
sources, and remains responsible for the paper's claims, citations, scripts,
scores, and conclusions.

\bibliographystyle{ACM-Reference-Format}
\bibliography{kpr-paper}

\end{document}